\documentclass[runningheads]{llncs}

\usepackage{orcidlink}
\usepackage{xcolor}
\usepackage[acronym]{glossaries}
\usepackage{hyperref}
\usepackage{csquotes}
\usepackage{listings}
\usepackage{tabularx}
\usepackage{ctable}
\usepackage{cleveref}
\usepackage{paralist}
\usepackage[colorinlistoftodos]{todonotes}
\usepackage{nameref}
\usepackage{enumitem,hyperref}
\usepackage{graphicx}
\graphicspath{ {./images/} }
\usepackage[T1]{fontenc}
\usepackage{tikz,lipsum,lmodern}
\usepackage[most]{tcolorbox}
\usepackage{float}
\usepackage[font=small,format=hang,parskip=1pt]{caption}
\usepackage[figureposition=bottom,tableposition=top]{caption}

\crefname{lstlisting}{listing}{listings}
\Crefname{lstlisting}{Listing}{Listings}

\begin{document}
\title{Designing NLP-based solutions for requirements variability management: experiences
from a design science study at Visma}
\titlerunning{Designing NLP-based solutions for requirements variability management}
%
%
\author{Parisa Elahidoost\inst{1,2}\orcidlink{0000-0003-4239-7838} \and
Michael Unterkalmsteiner\inst{1}\orcidlink{0000-0003-4118-0952} \and
Davide Fucci\inst{1}\orcidlink{0000-0002-0679-4361} \and 
Peter Liljenberg\inst{3}\and 
Jannik Fischbach\inst{2,4}\orcidlink{0000-0002-4361-6118}}
\authorrunning{P. Elahidoost, M. Unterkalmsteiner, D. Fucci, P. Liljenberg and J. Fischbach}
%
\institute{Software Engineering Research Lab SERL, Blekinge Institute of Technology, Valhallavägen 1, 37179 Karlskrona, Sweden \and
fortiss GmbH, Guerickestraße 25, 80805 Munich, Germany \and
Visma, Sambandsvägen 5, 35236 Växjö, Sweden\footnote{at the time of the research} \and
Netlight Consulting GmbH, Prannerstraße 4, 80333 Munich, Germany}
\maketitle              
%
\begin{abstract}

\textbf{Context and motivation}: In this industry-academia collaborative project, a team of researchers, supported by a software architect, business analyst, and test engineer explored the challenges of requirement variability in a large business software development company. \textbf{Question/ problem}: 
Following the design science paradigm, we studied the problem of requirements analysis and tracing in the context of contractual documents, with a specific focus on managing requirements variability. This paper reports on the lessons learned from that experience, highlighting the strategies and insights gained in the realm of requirements variability management.
\textbf{Principal ideas/results}: This experience report outlines the insights gained from applying design science in requirements engineering research in industry. We show and evaluate various strategies to tackle the issue of requirement variability.
\textbf{Contribution}: We report on the iterations and how the solution development evolved in parallel with problem understanding. From this process, we derive five key lessons learned to highlight the effectiveness of design science in exploring solutions for requirement variability in contract-based environments. 

\end{abstract}

\section{Introduction} \label{sec:intro}

Variability management is a key factor when considering the frequent necessity to modify and update software products. Variability signifies the capacity for alterations in both software products and models~\cite{10.1007/11763864_8}. Variability management includes activities such as representing variability in software artifacts explicitly throughout their lifecycle, managing interdependencies, and facilitating implementation. This management process is intricate and challenging, necessitating methodologies, techniques, and tools to be effectively executed~\cite{SCHMID2004259}.

In this study, Blekinge Institute of Technology (BTH) collaborated with Visma, a software company producing a payroll solution that supports hundreds of collective labour agreements (CLA). CLAs are contractual documents written in natural language that represent the basis for the development, testing, and configuration of the payroll system. CLAs within the same domain and related industries often share aspects relevant to implementation. Similarities across these documents usually represent a shared software configuration. A pain point for the company is identifying and managing such similarities, particularly considering that CLAs are renewed regularly. Currently, business analysts need to \textit{manually} sift through documents to assess similarity and maintain traceability between these shared configurations and CLAs. This task is labor-intensive, time-consuming, and error-prone.
The initial aim of the collaboration between BTH and Visma was to investigate the challenges related to the variability and traceability of configurations extracted from CLAs, and their impact on requirements analysis and testing. 
In this paper, we report our experience applying design science~\cite{design} to build, together with Visma, several iterative, Natural Language Processing (NLP) based solutions to address problems related to traceability when using CLAs as a source for requirements specifications. We report five lessons learned from this experience.
This paper presents insights for future industry-academia collaborations in the same area. Furthermore, it provides an example of the challenges that can arise when dealing with requirements extracted from regulatory documents, such as contracts.

The rest of the paper is organized as follows. Section~\ref{sec:RW} presents a brief overview of the existing literature on the application of NLP to traceability problems in software engineering and variability extraction from natural language documents. We detail our research goals and methodology in Section~\ref{sec:method}. In Section~\ref{sec:DS}, we report the results from five design iterations with associated lessons learned. Finally, Section~\ref{sec:con} concludes the paper.

\section{Related Work} \label{sec:RW}

A recent systematic review of 96 primary studies published between 2013 and 2021 on the application of NLP to traceability problems in software engineering~\cite{ZC23} highlighted the different syntax and semantics across artifacts as well as the difficulty to modelling tacit knowledge as the main challenges in this research area.
Michelon et al.~\cite{spl} addressed the challenges of maintaining, evolving, and composing variants of systems that evolve over space and time. They identify four key challenges in this area, focusing on the complexities introduced by the temporal and spatial evolution of system variants. Their work emphasizes the need for effective strategies to manage these evolving variants, aiming to improve the maintenance and evolution processes. 
Ferrari et al.~\cite{ferrari} developed tools, CMT and FDE, to facilitate the translation of requirements expressed in natural language into feature models, specifically feature diagrams. Their work aims to bridge the gap between the informal documentation of requirements and the structured representation needed for software product line engineering. These tools support the automated analysis of natural language documents, extracting relevant features, and structuring them into a formal model, thus streamlining the process of feature diagram creation and enhancing the accuracy and efficiency of software product development.
Reinhartz-Berger et al.~\cite{product} explored the use of semantic and ontological methods to analyze the variability within Software Product Lines (SPLs). Their work aims to understand and manage the complexity and diversity of SPLs by applying advanced information systems engineering techniques. Through their research, they propose a structured approach to capture and reason about the variability in SPLs, leveraging semantic and ontological considerations to improve the design and customization of software products. 
In regulatory requirements engineering, which is closely related to our study, NLP techniques have been employed to create traceability links between regulations and specifications. Jain et al.~\cite{JGS14} utilized five semantic similarity methods to connect use cases derived from specifications to regulatory documents, achieving a Medium Average Precision (MAP) of 0.72 in their evaluation of 69 insurance domain documents.
Sleimi et al.~\cite{SSS18} enhanced the systematic traceability of legal requirements by proposing a method to automatically extract metadata using domain-specific semantics. This approach was based on parsing rules developed from constituency and dependency parsing, refined with expert annotations of 200 legal statements in the traffic and vehicle regulation domain.
Li et al.~\cite{Yang17} provide a comprehensive and detailed systematic literature review of approaches and tools in the area of feature and variability extraction from natural language documents.
Besides showing that requirements and product descriptions are most commonly used in existing research, the authors show that several of the proposed approaches are neither accurate nor complete, limiting their practical use.
The authors present a pipeline, consisting of several NLP, information retrieval, and machine learning-based techniques.
In our research, we draw on NLP techniques related to paraphrase mining and plagiarism detection, as discussed by Pavlick et al.~\cite{Pavlick2015DomainSpecificPE}. They adapted a machine translation method to learn paraphrases from bilingual corpora, achieving a 4.2\% improvement over existing approaches with a 43.7\% Area Under the Curve (AUC) in their best model.
Ma et al.~\cite{Ma2019EssentiaMD} introduced an unsupervised technique for extracting paraphrases from small topic-specific datasets. Their method, based on word alignment to form a graph for generating paraphrase candidates, outperformed state-of-the-art approaches, significantly increasing recall by 247 to 460\% across various corpora while maintaining precision comparable to existing methods.
Folt\'ynek et al.~\cite{Foltynek2020} examined six word embedding models and five classifiers for identifying plagiarism, with their best classifier achieving 83.4\% to 99.0\% accuracy, surpassing human experts and current plagiarism detection systems
Wahle et al.~\cite{10.1007/978-3-030-96957-8_34} assessed five pre-trained word embedding models and eight neural language models on various texts, including paraphrased research papers and articles, with their top models outperforming human evaluators by an average of 2.5\%, achieving a 74.8\% to 80.5\% micro F1-score.
\section{Research Methodology} \label{sec:method}

In 2018, the Software Engineering Research Lab group at BTH initiated an eight-year research program\footnote{\url{rethought.se}} with initially nine companies from the telecom and finance domains. We had no prior collaboration experience with six of these companies. Hence, our focus was to understand their challenges in software engineering and match them with our expertise and interests. Visma, a large international accounting software company, was interested in our previous work on using test cases as requirements specifications~\cite{bjarnason2016multi}. In particular, they were interested in ``getting a mapping of requirements from laws and collective agreements and see how well they are covered by our test cases.'' In section~\ref{sub:I1}, we set out to address this research agenda. The collaboration included three champions~\cite{wohlin2011success} who acted as our contact points within Visma, a software architect (SA), a business analyst (BA), and a test engineer (TE). 
A significant challenge for business analysts is the manual identification and management of similarities between CLAs. This situation presents a clear research goal: 
\textit{to develop an automated solution that can efficiently and accurately identify and manage the similarities among CLAs, thereby reducing the reliance on manual processes, minimizing errors, and improving overall efficiency in the management of software requirements configurations.}

We followed the design science paradigm~\cite{design,Runeson2020} to design, implement, and evaluate a technical solution to fulfill our research goal. This paper reports on five iterations in which we develop, evaluate, and improve the solution design collaboratively with the three champions. The design science methodology encompasses three principal phases: initial \textit{problem conceptualization}, addressed in the first two iterations (Sections~\ref{sub:I1}-\ref{sub:I2}) through prototype development and cognitive task analysis; the \textit{solution design}, involving CLA analysis and the collection of feedback, primarily during iterations three to five (Sections~\ref{sub:I3}-\ref{sub:I5}); and a final phase of \textit{empirical evaluation}, where we have preliminary indications of the solution effectiveness. Each iteration is characterized by specific inputs, the actions undertaken, the resulting outputs, and the identified lessons learned.

\section{Design Iterations and Results} \label{sec:DS}


This section presents the design science iterations we conducted to understand the challenges of practitioners working with CLAs, build a set of candidate solutions to these challenges, and evaluate them. We report on the iterations and how the solution development evolved in parallel with problem understanding. Lessons learned synthesize the key takeaways from each iteration.  


\subsection{Iteration One} \label{sub:I1}
We scheduled an online meeting with SA, BA and TE to understand the problem better. We agreed to initiate a feasibility study to explore whether current NLP techniques can support BAs in evaluating to what extent test cases, which in the case of the company represent de-facto the requirements, cover statements in CLAs. 
SA provided us with CLAs and software feature configurations, which we used to implement a simple keyword extractor that matched terms between the two artifacts. 
\begin{figure}[htp]
    \centering
    \includegraphics[scale=0.55]{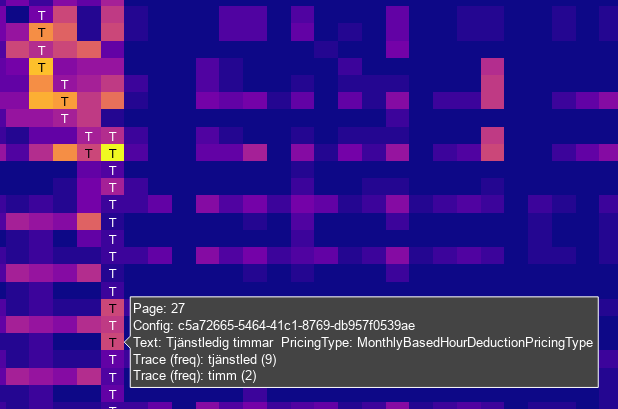}
    \caption{Visualization of term matching frequencies in feature configurations (rows) and pages in a CLA (columns). Ground truth marked with a ``T''.}
    \label{fig:heatmap}
\end{figure}
\raggedbottom
We visualize the terms matching in Figure~\ref{fig:heatmap}. Each row in the heatmap represents a feature configuration (43 in total), and each column represents a page (72 in total) from one CLA. The color in each cell represents the frequency of the terms (lower-cased, stemmed), extracted from the feature configuration, encountered on the page from the collective agreement. A cell marked with a $T$ signifies that this page is the source for a feature configuration---i.e., the ground truth. We used this visualization to discuss with Visma the feature configurations in the areas of the heatmap where the approach worked well (upper left) and where it did not (lower right). We also discussed using the heatmap to discover areas in the CLA that contain potentially relevant information (upper right area) or to guide the BA in reading unseen collective agreements.
Although we received positive feedback from SA, BA and TE, the heatmap did not solve BA's and TE's specific problems when working with CLAs, which they could not articulate then. This misalignment between problem and candidate solution highlighted other challenges the BA and TE encountered when analyzing the CLAs, which the researchers or practitioners had not fully understood yet.

\begin{tcolorbox}[colback=black!5!white,colframe=black!75!black,title=Lesson learned:]
  Even though the initial meeting with the SA, BA, and TE produced an actionable problem description that could be solved quickly through a proof-of-concept prototype, we did not yet have an in-depth problem understanding. We realized that articulating these challenges in a vacuum, without illustrating the tasks these stakeholders struggle with, is ineffective. However, the heatmap helped researchers and practitioners realize that our understanding of the problem was not aligned. Concrete artifacts created early in a research project can catalyze the discovery of such gaps. 
\end{tcolorbox}


\subsection{Iteration Two} \label{sub:I2}
We used Applied Cognitive Task Analysis (ACTA)~\cite{militello1998applied} to elicit a detailed description of how the BA and TE perform a particular task, described below, involving the CLAs. We designed a simple data collection instrument in a spreadsheet, capturing the performed sub-tasks, the knowledge needed to perform them, and the associated challenging cognitive activities. Our goal was to understand which tasks require cognitive skills that, with the increase of supported CLA, can be difficult to perform. Such tasks are candidates for designing and evaluating a technical solution. As a basis for the analysis, we identified together with BA and TE the following tasks: (1) develop, based on a CLA, a product configuration for salary payments and adapt, if necessary, the existing product configuration structure (BA); (2) assemble, based on a collective agreement, a test suite and create, if necessary, new test cases (TE). Based on the tasks they are performing, BA and TE have related, yet slightly different, challenges using the CLAs. 
The variation points (i.e., where the agreement can vary) and the variants (i.e., the options at a particular variation point) of CLAs regarding salary payment were not yet completely defined in the company. Hence, when the BA analyzes a new agreement, a change in the configuration may be necessary. When a new agreement is analyzed, it is tedious and error-prone to manually scan the agreements and compare them to find new variants.
The decision to test a new agreement with an existing test suite or whether to change or add a test case relies on TE ability to recall the contents of CLAs. Wrong assumptions can lead to incomplete or wrong test cases. Furthermore, identifying missing test cases is difficult as there is currently no structured way to analyze and measure test coverage for a CLA.

\begin{tcolorbox}[colback=black!5!white,colframe=black!75!black,title=Lesson learned:]
When we dissected the task performed by BA and TE into individual steps, we understood \emph{why} their approach, while effective on a small number of agreements, did not scale. The identified problem is certainly actionable---i.e., we can devise several solutions, built on existing research, that can support them in analysing hundreds of agreements. We can also evaluate the effectiveness of the solutions by comparing them to a baseline, contributing evidence to the research area. Making the thought process of experts explicit and uncovering the atomic actions they perform during analysis and their connections are powerful ways to strip away the uncertainty surrounding tasks that stakeholders perceive as challenging. Accordingly, ACTA can be a useful tool to gain a common problem understanding.
\end{tcolorbox}

\subsection{Iteration Three} \label{sub:I3}
\begin{figure}[h!]
    \centering
    \includegraphics[scale=0.28]{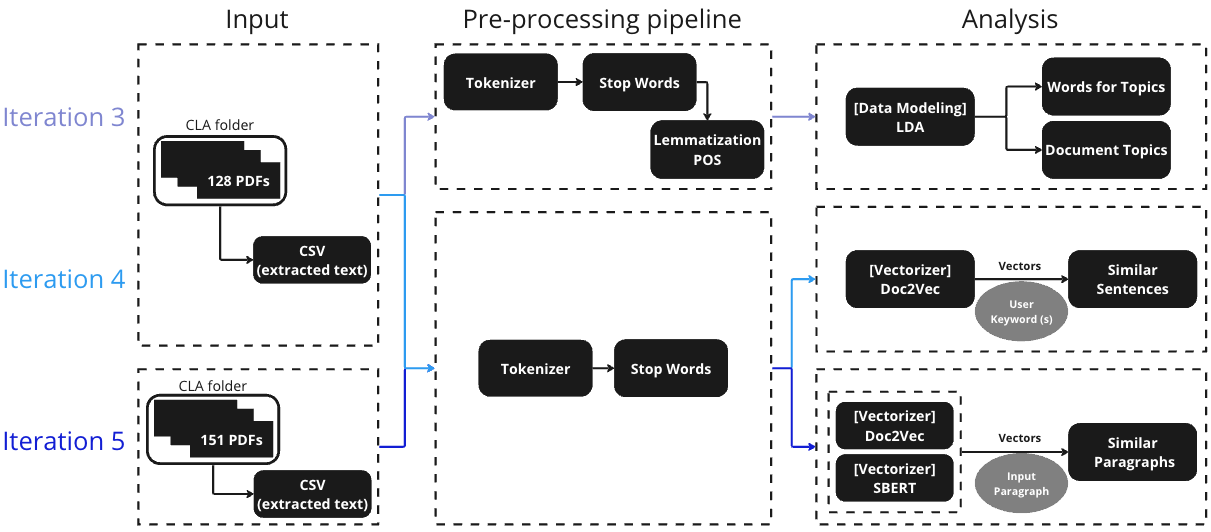}
    \caption{Graphical depiction of the proposed method - Iterations three, four and five}
    \label{fig:analysis}
\end{figure}

We used the results from ACTA to refine the problem statement. Accordingly, our new goal was to identify common variants, allowing BA to analyze one aspect that spans all CLAs. Fulfilling this goal allows the BA to validate the configuration developed based on a few agreements against the remaining CLAs. To devise a technical solution to the problem, we obtained 128 CLAs \footnote{Contracts are available upon request \href{https://zenodo.org/records/10640865}{https://zenodo.org/records/10640865}}, written in the Swedish language, and a list of 17 keywords that the BA used to search specific passages in the CLAs. 
Our proposed solution for this challenge employs topic modeling and information filtering to identify key topics and significant themes in each document. Topic modeling is particularly well-suited for this problem as it excels in uncovering hidden variants and themes in large text collections by detecting groups of frequently co-occurring words. In our analysis, we employed the Latent Dirichlet Allocation (LDA) technique, which views documents as compositions of various topics identified by the LDA algorithm. This aligns closely with our aim to efficiently filter and categorize document content.
We use Figure~\ref{fig:analysis} to illustrate the variations in input, pre-processing, and analysis we introduced in the remaining iterations of solution development. In iteration three, we started by pre-processing the text that had been extracted from 128 CLAs. We extracted paragraphs and their associated document name and page number. We removed stop words, lemmatized the paragraph text, and applied part-of-speech (POS) tagging.


\raggedbottom



To identify the number of topics in all documents, we calculated the coherence score---i.e., a metric for measuring the semantic interpretability of the top terms~\cite{OCALLAGHAN20155645}. We represented topics with the top-$N$ tokens with the highest likelihood of belonging to a specific topic. The coherence score quantifies the extent to which these words exhibit similarity to one another. In our analysis, 23 was the best number of coherent topics. We used the BA's input keywords to validate to ascertain whether the identified topics effectively captured these keywords. This approach aids BA in determining which documents encompass the specified keywords by locating them within the topics.
We shared our results in a meeting with the three champions. 


While BA and TE found the results interesting, there were reservations about their practicality and effectiveness in simplifying their tasks. The feedback from our discussion indicated that topic modeling might not adequately address the problem. Instead, a more detailed analysis that pinpoints the exact page and location of keywords in the documents was deemed more beneficial. Therefore, the next iteration focuses on providing these specific, granular details to better meet the champions' needs.

\begin{tcolorbox}[colback=black!5!white,colframe=black!75!black,title=Lesson learned:]
To ensure that identifying variability in CLAs produces practical and actionable insights, the variants and keywords we identify must be easily traceable. This involves pinpointing the exact location of these variants, down to the specific sentence on a designated page in a CLA. Such precision bridges the gap between general observations and detailed, applicable data, which is essential for effectively managing variability in CLAs. Additionally, conducting meetings that include all the beneficiary roles (BA, TE, and SA) simultaneously has proven beneficial. These interdisciplinary discussions enhanced the understanding of each role's responsibilities and improved the collective workflow.
\end{tcolorbox}

\subsection{Iteration Four} \label{sub:I4}

This iteration aims to pinpoint the exact page and sentence associated with each keyword. This precision ensures useful traceability and enriches the context, empowering BA to determine the pertinence of every finding. To this end, we explored the possibility of utilizing text representation approaches like word embeddings. 
In our analysis, we used Doc2Vec as we wanted to identify similar sentences. Doc2Vec generates distributed representations of variable-length pieces of text, such as sentences, paragraphs, or entire documents. 
During this iteration, conducted in 2021, the availability of models for less common languages like Swedish was limited. Therefore, we trained a Doc2Vec model based on the Swedish Wikipedia. 
The outline of the proposed method for iteration four can be seen in Figure~\ref{fig:analysis}. After pre-processing, we used the pre-trained Doc2Vec for vectorizing all the input documents and keywords and calculated the cosine similarity between them. 
We distributed a spreadsheet to BA and TE featuring a list of keywords, each accompanied by relevant paragraphs from the CLAs and page numbers of semantically similar paragraphs, to assess our findings. We recognized some difficulty in understanding how a spreadsheet could be beneficial for daily tasks. We created a prototype user interface to display and filter the contents, tailoring it to the BAs' intended use cases. This prototype served as a practical demonstration of the tool potential benefits, making it easier for BA to see its applicability and usefulness in their daily work.

\begin{figure}[htp]
    \centering
    \includegraphics[scale=0.3]{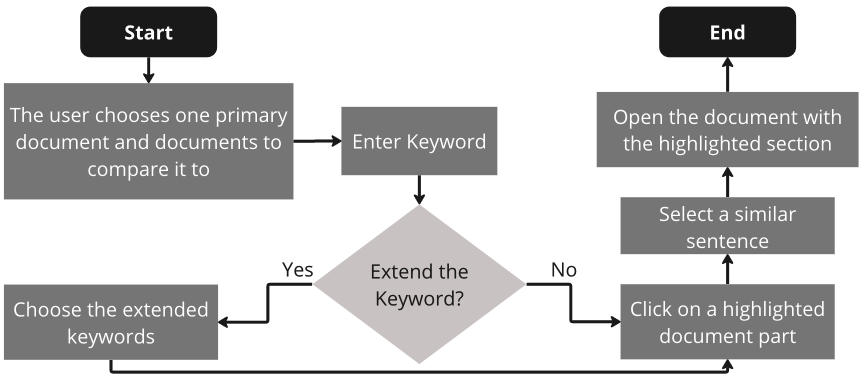}
    \caption{User task flow diagram for CLA analysis}
    \label{fig:flow}
\end{figure}

Figure~\ref{fig:flow} shows the user task flow diagram for the CLA analysis interface mockup. 
\raggedbottom
We showcased the prototype during a workshop. The prototype displayed the key elements of the final application design and aided the analysts in better articulating their requirements. 
The proposed direction for the forthcoming iteration is to gather additional insights from the analysts to further refine the prototype. The desired functionality is to retrieve all semantically similar paragraphs from various CLAs given an input paragraph. We aim to enrich the context by extracting paragraphs before and after the target text, highlighting differences from the input to streamline the analysis.

\begin{tcolorbox}[colback=black!5!white,colframe=black!75!black,title=Lesson learned:]
Prototyping emerged as a crucial factor in the latter stages, significantly influencing the integration of user-specific preferences and requirements into the core structure of our final research artifact. This strategy markedly augmented the engagement and interaction levels among our champions, showcasing how prototyping serves as an essential tool in understanding the research outcome. It emphasizes the transition from preliminary problem understanding to a more refined, user-centered research phase, ensuring that we not only addresses the identified issues but also aligns closely with the users' practical needs and expectations.
\end{tcolorbox}

\subsection{Iteration Five} \label{sub:I5}

\begin{figure}[htp!]
    \centering
    \includegraphics[scale=0.4]{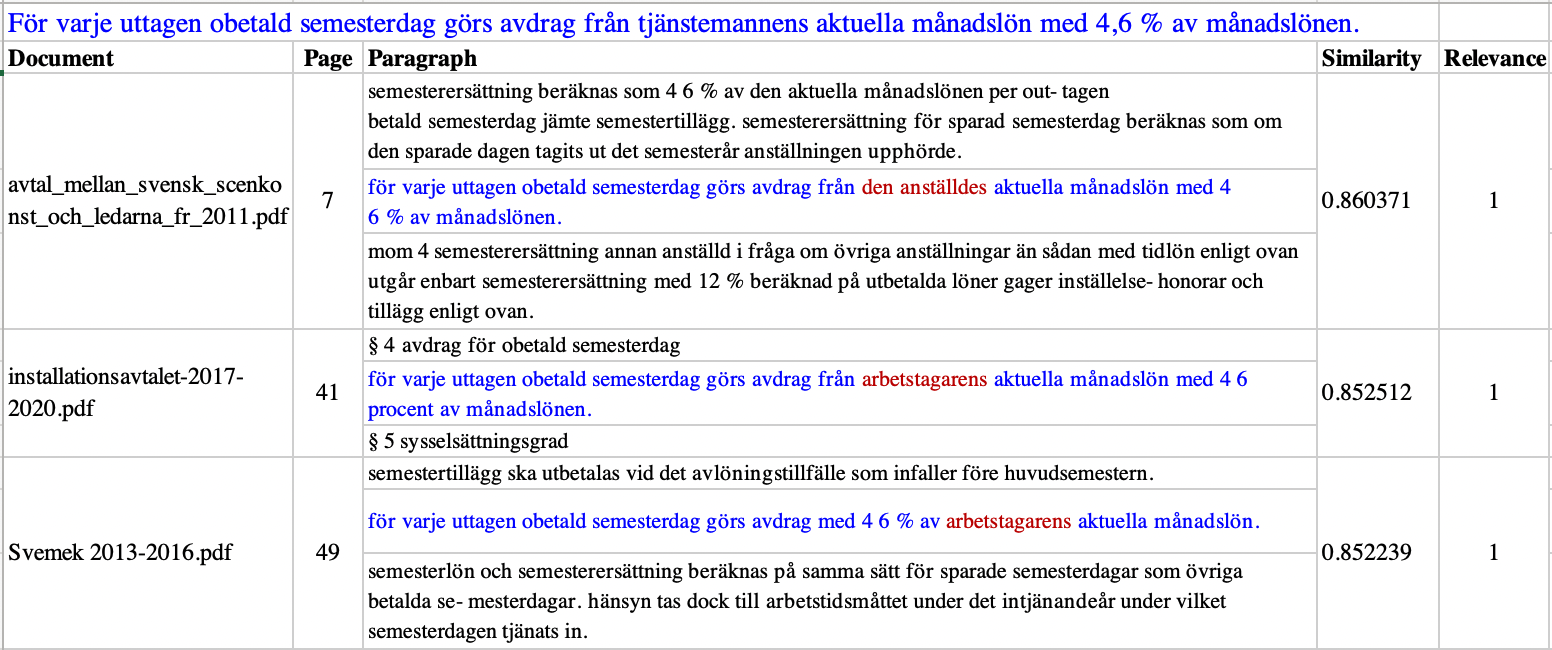}
    \caption{Snippet of the result}
   \label{fig:final}
\end{figure}

In this iteration, the analysis goal is to identify all the paragraphs of all the CLAs that are similar to the given paragraph and, therefore, likely associated with the same keyword. This allows the BA to know which CLAs are not covered by the current manually created workflow for each keyword. The BA can also analyze the newly identified paragraphs to identify variations that are not yet considered in the current configuration model. Moreover, for each keyword, we identify all shared variants relating to that keyword. 
We received 14 keywords, each with a variant sentence from a CLA, including page numbers, to identify similar variants across all CLAs for software configuration. For example, in the sentence \textit{``The sick leave deduction must be 20\% of the average sick pay per week.''} the keyword is \textit{sick leave deduction} and \textit{20\% of the average sick pay per week} is the variant for configuration. For this iteration, we received a total of 151 CLAs. Figure~\ref{fig:analysis} presents the outline of the proposed method for iteration five. 
Unlike the previous iteration, the input for similarity calculation is a sentence that contains the keyword and the configuration variant. The Doc2Vec model has been trained on the Swedish Wikipedia and the existing CLAs. We determine the similarity between paragraphs by computing their cosine similarity and then ranking them in descending order. We add context---i.e., previous and next paragraphs---to the selected ones. 
We also experimented with \textbf{SBERT Semantic Search} which generates embeddings that have multiple vector representations of the same word based on its context. Thus, BERT embeddings are context-dependent. There exist several pre-trained models
for different purposes like Semantic Textual Similarity, Clustering, and Paraphrase mining~\cite{reimers-2019-sentence-bert}.
We used SBERT for vectorizing sentences and paragraphs and then used the community detection function (which calculates cosine similarity) to list the similar paragraphs to the input paragraph. Additionally, we highlighted differences between each input paragraph and its similar counterparts for a clearer presentation.
Figure~\ref{fig:final} presents a sample output from the Doc2Vec model corresponding to the keyword input paragraph ``Holiday deduction'' \footnote{Swedish: \textit{Semesteravdrag}} and the sentence ``For each unpaid vacation day taken, a deduction of 4.6 \% of the monthly salary is made from the employee's current monthly salary.'' \footnote{Swedish: \textit{För varje uttagen obetald semesterdag görs avdrag från tjänstemannens aktuella manadslon med 4,6 \% av månadslönen.}}. We processed 10 paragraphs, each matched to a keyword, using both the Doc2Vec and SBERT models, and presented the top-20 results from each to the SA for review. They evaluated the results marking relevant and non-relevant matches. 
\raggedbottom
To assess both models, we computed their average precision (AvP)---i.e., a metric condensing the precision-recall relationship into a single value~\cite{Zhang2012StatisticalIO}. 



Table~\ref{tab:AP} presents a comparative summary of the outcomes. Doc2Vec achieved marginally superior results compared to SBERT. Notably, both models yielded better performance with inputs that included formulas and numbers, as opposed to the more brief, text-only inputs like paragraph 4.
\begin{table}[htp]
\centering
\scriptsize
\caption{Average precision results for Doc2Vec and SBERT}\label{tab:AP}
	\begin{tabular}{p{4cm}p{3cm}p{3cm}l}\toprule  
		Input paragraphs &  Doc2Vec AveP   &  SBERT AveP  \\\midrule  
            paragraph 1       & 1 & 0.8987\\
            paragraph 2       & 0.6044 & 0.56\\
            paragraph 3       & 0 & 0  \\
            paragraph 4       & 0 & 0  \\
            paragraph 5       & 1 & 1\\
            paragraph 6       & 0.8571 & 0.7656 \\
            paragraph 7       & 0.9601 & 0.8807\\
            paragraph 8       & 1 & 0.8 \\
            paragraph 9       & 1 & 1 \\
            paragraph 10      & 1 & 0.887  \\\midrule 
            Mean Average Precision    &  0.74216  &  0.679 \\\bottomrule
	\end{tabular}
\end{table}

Based on the insights of iteration four, we developed a new user interface prototype that embodied our latest analysis, offering a clearer visualization of the use case. During the concluding meeting, upon reviewing the prototype, BA acknowledged that the analytical findings are be beneficial for their routine work. 

\begin{tcolorbox}[colback=black!5!white,colframe=black!75!black,title=Lesson learned:]
We recognized that automating the identification of shared variants cannot be fully achieved. When it comes to legal or contractual analysis, expert input from those with domain knowledge remains essential, especially when considering the context of the input.
The iterations helped the champions refine their needs and objectives. Design Science enabled continuous improvement in design, enhancing both the user interface and user experience through iterative refinements. Each phase is carefully enhanced, drawing on user feedback and suggestions. Implementing this iterative methodology proves beneficial in industry-academia partnerships, particularly when there is ambiguity about the problem context for both collaborators.
\end{tcolorbox}
\section{Conclusion} \label{sec:con}

The collaboration between the BTH and Visma provided a context for investigating the complexities of requirement traceability in the presence of high variability, typical of contractual documents. In this work, we showed the efficacy of design science methodology in addressing the industry-relevant challenge of managing variability and maintaining traceability among CLAs. The lessons learned from this effort underscore the potential benefits of applying design science to gain problem understanding. This is further illustrated through the iterative nature of our solution development, which evolved with our understanding of the problem. Throughout this experience, one of the significant findings was the realization that the goal of completely automating the identification of shared configurations is not entirely feasible. Our results underscore the need for expert intervention, particularly in tasks involving legal or contractual analysis. Nevertheless, the stakeholders involved in our study acknowledged that even a modest degree of automation can be beneficial. 
These findings, while specific to this context, hint at broader implications for similar challenges in other industries, suggesting a potential for generalizing the approach to address variability and traceability in different domains.
Our insights contribute to both academic research and industry practices, offering tangible examples of the hurdles faced when managing contract-based requirements. 

\section*{Acknowledgment}
We would like to thank all employees at Visma who supported our study. This work was further supported by the KKS foundation through the S.E.R.T. Research Profile project at Blekinge Institute of Technology.

%
%
%
%

\bibliographystyle{splncs04}
\bibliography{bibliography}






\appendix

\end{document}